\documentclass[reprint,aps,ams,amsmath,prl,twocolumn,longbibliography,superscriptaddress]{revtex4-1}
\usepackage{graphics}
\usepackage{graphicx}
\usepackage{epsfig}
\usepackage{amsmath}
\usepackage{amssymb}
\usepackage{amsfonts}
\usepackage{bm}
\usepackage{color}
\usepackage{bbm}
\usepackage[T1]{fontenc}
\usepackage[bookmarks=false,linkcolor=blue,urlcolor=blue,colorlinks,citecolor=blue]{hyperref}
\usepackage{libertine}

\begin{document}

\title{Electron liquids in solids: hydrodynamic description inspired by ideas of Radii Gurzhi}

\author{A. N. Kalinenko}
\affiliation{B. Verkin Institute for Low Temperature Physics \& Engineering, NAS of Ukraine, Kharkiv 61103, Ukraine}

\author{A. Levchenko}
\affiliation{Department of Physics, University of Wisconsin-Madison, Madison, Wisconsin 53706, USA}

\date{November 8, 2023}

\begin{abstract}
This is the preface article written for the special issue of Low Temperature Physics on the captivating topic of electron hydrodynamics dedicated to the pioneering work of Radii Gurzhi. 
The article features a brief synopsis of Gurzhi's seminal contributions that have played a pivotal role in shaping this continuously evolving area of condensed matter physics. 
This tribute is followed by a brief introduction to the collection of contributed papers published in the issue representing recent research in this dynamic field.       
\end{abstract}

\maketitle

Radii Mykolayovych Gurzhi (11.08.1930–03.08.2011) was a renowned scientist, and the author of a series of outstanding results and effects in solid-state physics, particularly in the field of the theory of kinetic phenomena in metals and dielectrics. 

In 1952, after graduating from Kharkiv State University, R. M. Gurzhi began his scientific career as a graduate student at the Lebedev Physical Institute of the USSR Academy of Sciences. Starting from 1957, he worked at the Kharkiv Physics and Technology Institute in the theoretical department under I. M. Lifshitz. In 1974, Radii Mykolayovych transferred to the Low Temperature Physics Institute of the National Academy of Sciences of Ukraine, where he led the Department of electronic properties of solids for nearly 30 years. Throughout these years, he made great efforts to preserve and advance the best traditions of the Kharkiv school of theoretical physics, which combines a highly professional mathematical approach with a clear physical interpretation of phenomena. From 2003 to 2011, he worked as the chief research scientist in the unified theoretical physics department.

Radii M. Gurzhi’s contribution to the development of modern theory of transport phenomena received wide recognition among both theorists and experimentalists. Many of his results became a significant contribution to the development of a modern understanding of charge and energy transport mechanisms in solids and found their way into well-known textbooks and monographs, which have already become classics. 

Notably, his work included the development of the theory of high-frequency properties of metals in the infrared region of the spectrum, the derivation of the quantum kinetic equation, the theory of hydrodynamic effects in metals (the Gurzhi effect), dielectrics, and semiconductors, and the theory of low-temperature electrical conductivity in pure metals. In the field of low-dimensional conductors, the series of works by R. M. Gurzhi and his students on low-temperature transport in two-dimensional systems became a novel contribution. For these works, he was awarded the State Prize of Ukraine in Science and Technology in 1997. Starting from the early 2000s, Radii M. Gurzhi actively worked in a new scientific area for him -- spintronics, where, together with his students, he predicted a series of spin dynamics effects in magnetically nonuniform conducting systems, which could serve as the basis for creating new spintronic devices. Regarding the scientific achievements of R. M. Gurzhi, the following can be distinguished.

\begin{figure}[t!]
 \centering
 \includegraphics[width=0.7\linewidth]{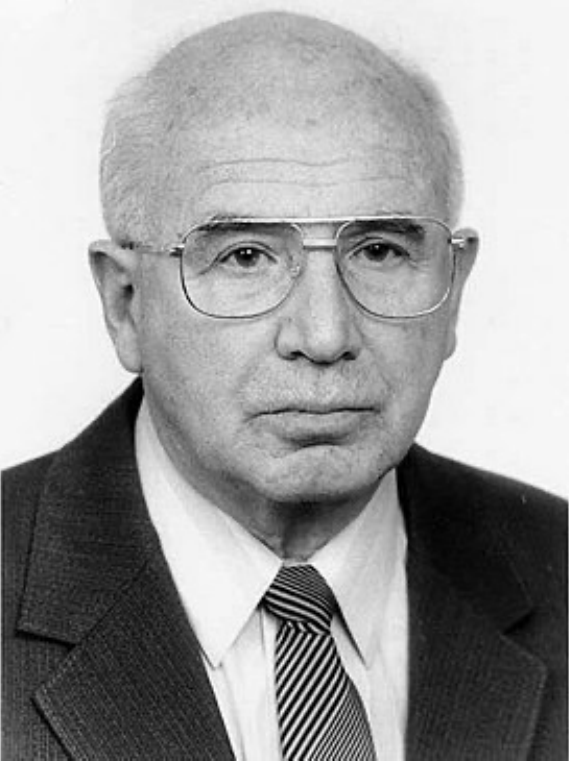}
 \caption{Radii Mykolayovych Gurzhi (11.08.1930–03.08.2011)}
 \end{figure}

\paragraph{1. Quantum phenomena in the kinetics of solids.} Radii M. Gurzhi proposed a method for deriving a quantum kinetic equation \cite{Gurzhi-JETP58a}, which was a development of M. M. Bogolyubov’s ideas regarding quantized systems. (There is a separate section in K. P. Gurov’s monograph titled "Derivation of Gurzhi’s Quantum Kinetic Equation"). He constructed a theory of high-frequency properties of metals in the infrared region of the spectrum \cite{Gurzhi-JETP58b,Gurzhi-JETP59}, which stimulated many new experimental studies. He derived a kinetic equation that takes into account the uncertainty of phonon energies that scatter. Based on this, a solution to the Pomeranchuk problem in the thermal conductivity of dielectrics was provided \cite{Gurzhi-JETP64}. For the first time at the microscopic level, the famous Landau–Lifshitz equation with an explicit expression for the relaxation term was derived \cite{Gurzhi-JETP67}. The method has been repeatedly used by other authors to consider non-standard kinetic problems. 

\paragraph{2. Hydrodynamic effects in 3D structures.} These works are dedicated to the study of kinetic phenomena in which normal collisions between quasiparticles play a fundamentally important role. It has been shown that normal collisions, by themselves, do not lead to the emergence of resistance (electrical or thermal), but they can qualitatively change the results of other scattering processes. As a result, both the order of magnitude of kinetic coefficients and their dependencies on characteristic parameters change.  

First, it is worth mentioning the Poiseuille flow of phonon gas in dielectrics \cite{Gurzhi-JETP64}, which was experimentally discovered "fairly quickly" when a multiple increase in the thermal conductivity coefficient (up to 10 times in the best $^{4}$He monocrystals) and a significant change in its temperature and size behavior were observed, as predicted by theory.

Another extraordinary effect that has been dubbed the Gurzhi effect in the literature is the prediction of the existence of a "hydrodynamic minimum" in the temperature dependence of the electrical resistance of metals \cite{Gurzhi-JETP63}. This phenomenon occurs when, under certain conditions, frequent normal collisions significantly slow down the transport of non-equilibrium momentum to the surface of the sample, where relaxation takes place. Thus, as the temperature increases, the resistance of such a sample decreases, rather than increasing as is typically the case with metals. This effect, although physically striking and beautiful, turned out to be quite challenging to experimentally observe. Predicted as far back as 1963, it awaited experimental confirmation for many decades.

Among other results, we can mention predictions regarding the role of high-order anharmonicity in the thermal conductivity of ferroelectrics, non-local hydrodynamics of phonon gas in dielectrics (both of these effects have been observed in experiments), and unexpected properties of the second sound in compensated metals, dielectrics, and semiconductors \cite{Gurzhi-SPU68}. It turned out that the structure of the second sound wave and its propagation velocity depend significantly on the nature of quasiparticles and the mechanisms of their interaction.

\paragraph{3. Low-temperature electrical conductivity of pure metals.} R. M. Gurzhi, together with O. I. Kopeliovich, formulated a method that allowed, from a unified perspective in terms of non-local electron diffusion on the Fermi surface, to describe a wide range of kinetic phenomena in pure metals at low temperatures \cite{Gurzhi-JETP72}. This enabled the prediction of a whole series of qualitatively new effects, characterized by the close interplay of electron dynamics on the Fermi surface with specific properties of electron-phonon interaction. The results are well-known and have been included in the book "Physical Kinetics" by E. M. Lifshitz and L. P. Pitaevskii, where a section titled "Electron Diffusion on the Fermi Surface" is dedicated to them.

\paragraph{4. Kinetic phenomena in low-dimensional conducting systems.} R. M. Gurzhi, together with his students, demonstrated that in systems with two-dimensional conductivity, instead of a single conservation law for the total momentum of phonons and electrons, an infinite number of conservation laws arises in a certain approximation \cite{Gurzhi-JETP82}. This served as a starting point for a significant reconstruction of the theory of kinetic phenomena in layered conductors such as graphite acceptor compounds or transition metal dichalcogenides. Fundamentally new mechanisms of momentum relaxation were predicted by R. M. Gurzhi and his students in two-dimensional degenerate electron gases that interact with each other \cite{Gurzhi-Adv87}. The absence of scattering processes leads to significantly different effects of frequent normal electron-electron collisions on the evolution of electron distributions with different symmetries. As a result, the evolution of the antisymmetric momentum distribution (which corresponds to "current" states) occurs much more slowly than in the 3D case. This is a key factor leading to a series of unconventional effects in the dynamics of low-dimensional electron gases, such as the evolution of two-dimensional electron beams. These effects have been experimentally observed in 2D systems based on heterojunctions.

\paragraph{5. Low-dimensional hydrodynamics.} Another extraordinary consequence of the peculiarities of electron collision mechanisms in a 2D electron gas (2DEG) is the fundamentally new "one-dimensional" hydrodynamics of a two-dimensional electron gas. This was predicted by R. M. Gurzhi and his students in works from the late 1980s to the mid-1990s \cite{Gurzhi-JETP86,Gurzhi-JETP89,Gurzhi-PRL95}. Unlike the 3D case, the symmetry of the nonequilibrium electron distribution with respect to momentum becomes crucial here. The influence of collisions is significantly different: symmetric distributions evolve rapidly, while asymmetric ones (including current distributions) evolve very slowly. As a result of collisions, constant processes of "electron-hole" conversion occur, effectively leading to a regime of "one-dimensional" diffusion of the nonequilibrium momentum carrier. In confined geometries, this process continues until the momentum is lost due to dissipative processes or diffusive scattering at the boundary of the 2DEG channel. Manifestations of such dynamics in electronic flows have been experimentally observed.

\paragraph{6. Spin dynamics effects in magnetically nonuniform conducting systems.} R. M. Gurzhi's group proposed the concept of an electrically controllable spin transistor based on the spatial distribution of spin components of an electric current in a hybrid spatially magnetically nonuniform structure. They also investigated the dynamics of spin-polarized electron fluid in conditions where normal electronic collisions prevail and the Poynting flow regime is realized. It was demonstrated that the mechanisms of spin transport undergo qualitative changes when electron-electron scattering of current carriers prevails due to the effect of mutual "friction" between electrons of different spins \cite{Gurzhi-PRB00,Gurzhi-PRB03,Gurzhi-PRB06}.

Overall, it should be noted that the theoretical scientific results of Radii Mykolayovych Gurzhi and his students sometimes significantly preceded the experimental possibilities of their time. Eventually, numerous beautiful and highly physical effects predicted long before the 21st century find realization in modern experiments and capture the attention of contemporary researchers.

The authors of the papers in the present collection were originally invited by the guest editors. Submitted papers then underwent peer review by the expert referees following the standards of the journal Low-Temperature Physics to ensure both high quality and relevance of the topics. We are delighted with the outcome as the collection of papers in this Special Issue covers many topics where Gurzhi made original pioneering contributions. In particular, these are the topics of electron hydrodynamics and kinetics in low-dimensional systems. 

For example, the paper by Edvin G. Idrisov \textit{et al.} \cite{Idrisov} addresses the role of spin-orbit coupling on the transport properties of the two-dimensional electron system in the hydrodynamic regime. For this purpose, the authors use the semiclassical Boltzmann equation and projection into the slow modes, which is, in essence, an approach initially applied by Gurzhi in the context of his work on “resistance minimum”. A generalized Navier–Stokes equation is then derived, and it is shown how spin-orbit effects influence the viscosity of electron liquid and the enthalpy of the system. 

The nonclassical hydrodynamics in 2D electron fluids, which includes several viscous modes with non-Newtonian viscosity, is elaborated in the work of Serhii Kryhin and Leonid Levitov \cite{Kryhin}. This transport regime was overlooked in the previous works and its physical origin can be rooted in Gurzhi’s prediction that different modes of the electron distribution function decay with the parametrically different time scales. 

In a parallel vein, it is shown in the work of P. Pyshkin and A. Yanovsky \cite{Pyshkin} that the difference between the relaxation times of the antisymmetric and symmetric momentum distributions in electron collisions may lead to a peculiar contact hydrodynamic-like effect. Namely, the appearance of a temperature-dependent potential difference and, consequently, resistance in the four-terminal measurements, which is distinct from the Sharvin–Landauer-like contact resistance.

The review by D. B. Gutman \textit{et al.} \cite{Gutman} provides an in-depth discussion of thermal conductivity in a one-dimensional electron liquid going beyond the parading of the Luttinger liquid theory. This paper covers the multi-mode hydrodynamic theory that is obtained by projecting the fermionic kinetic equation on the zero modes of its collision integral. It is then shown that the interaction between hydrodynamic modes leads to renormalization and consequently to anomalous scaling of the transport coefficients. 

The work of Graham Baker \textit{et al.} \cite{Baker} addresses the nonlocal electrical transport in anisotropic metals with an arbitrary Fermi surface. A simple but accurate model of a collision operator in the Boltzmann equation with a finite number of quasi-conserved quantities is used to derive closed expressions for the wave vector-dependent conductivity. This approach enables to investigate the interplay of momentum-conserving and momentum-relaxing scattering processes on the nonlocal response. 

The role of collective modes in electron hydrodynamics is investigated in the paper by Dmitry Zverevich and Alex Levchenko \cite{Zverevich}. While viscosity and thermal conductivity renormalizations originating from plasmon fluctuations are shown to be weak, plasmon resonances in the electron double-layer systems are shown to play the dominant role in the drag resistance. Interestingly, in systems without Galilean invariance, fluctuation-driven contributions to dissipative coefficients can be described only in terms of hydrodynamic quantities: intrinsic conductivity, viscosity, and plasmon dispersion relation. 

The effect of disorder on the energy transfer induced by Coulomb interaction in electron bilayers is investigated in the paper by Alex Kamenev \cite{Kamenev}. It is shown that, for the sufficiently closely spaced layers, the elastic scattering of electrons on impurities, leading to the diffusive spreading of density fluctuations, combined with the dynamical screening of the Coulomb interaction, leads to a dramatic increase in the thermal transfer conductance surpassing the radiative mechanism defined by the Stefan-Boltzmann law. 

\paragraph{Acknowledgment.} This translation of the original article in Ukrainian \cite{Kalinenko} was prepared during the program "Quantum Materials With and Without Quasiparticles" held at the  
Kavli Institute for Theoretical Physics, Santa Barbara, supported by the National Science Foundation under Grants No. NSF PHY-1748958 and PHY-2309135. 

\paragraph{Note.} The entire published FNT special issue on the topic of electron hydrodynamics can be found by following the link \href{https://fnt.ilt.kharkov.ua/index.php/fnt/issue/view/i49-12}{FNT Vol. 49 No. 12 (2023)}. 


\bibliography{Gurzhi-biblio}

\end{document}